# DEVELOPMENT OF DETECTOR ACTIVE ELEMENT BASED ON THGEM

V. Inshakov, V. Kryshkin, V. Skvortsov, A. Sytin

(*Institute for high energy physics, Protvino*)

N. Kuzmin, S. Sychkov

(*Joint institute for nuclear research, Dubna*)

Abstract

For radiation-hard, high counting rate detectors (hadron calorimeters, tracking detectors etc.) as an active element we considered a thick gas electron multiplier. There was carried out technological and design study to optimize the element structure. There are presented the measurement results and the next plans.

## 1. Introduction

The proposed increase of LHC luminosity by order of magnitude [1] posed a problem to upgrade the CMS active elements of the endcap hadron calorimeters designed and built at IHEP [2]. There were analyzed the requirements to the performance of the hadron calorimeters at high luminosity, considered different options of detectors and suggested a solution. We summarize here the activity in this field and present the results of the measurements and the plans.

The obvious requirements to detectors operating at harsh radiation environment are the following ones:

- high radiation resistance (~60 Mrad);
- fast signal response;
- high rate capability.

Besides the detector must be simple (cheep), reliable (access to the apparatus is very limited) and robust (the assemblage of the detectors will be at height of 6 stores building). We came to a conclusion that the thick gas electron multiplier (THGEM) [3] is the most closely complying with these demands. During the last years there were a big activity in this field and characteristics of THGEM were systematically studied. So we were guided by these results and had strictly practical goal – to optimize the chamber design for production technology.

## 2. Experimental setup

Fig. 1 shows the design of THGEM that differs from GEM by only an order of magnitude larger dimensions of multiplying electrode – thickness, hole diameter, hole pitch. Large dimensions allow to use for production the standard equipment for PCB manufacturing.

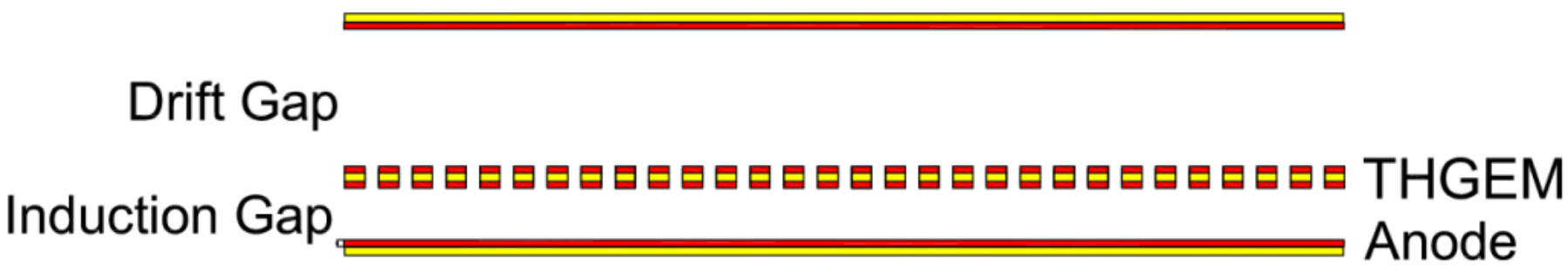

Fig. 1. Schematic view of the THGEM.

Fig. 2 shows the THGEM electrode used for the measurements. There are rims around the holes to reduce the probability of gas breakdown. The hole diameter is close to the thickness and the hole pitch is 800 μm. These electrodes were manufactured at IHEP workshop using CERN technology [4]. After production electrodes were cleaned in ultra-sound bath, holes were hydro abrasively cleaned and dried.

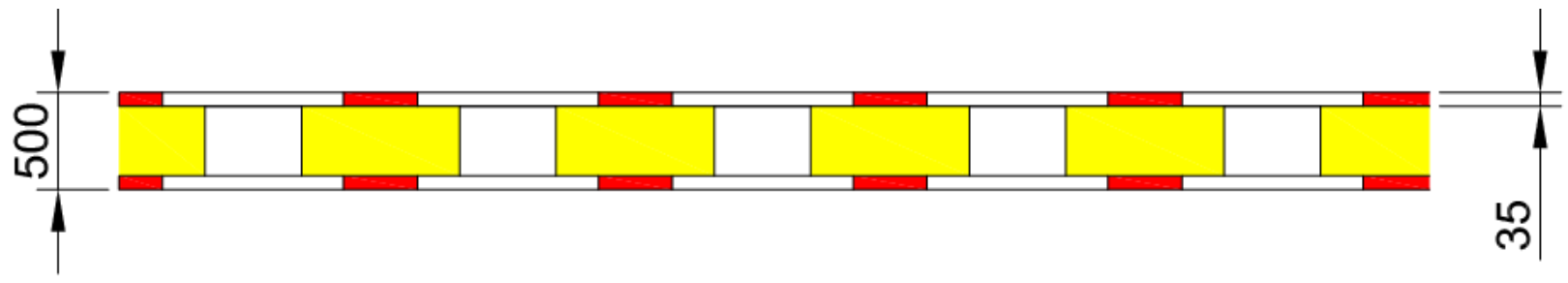

Fig. 2. The cross section of the THGEM – PCB 500 μm thick covered with 35 μm copper from both sides.

Fig. 3 shows the microscope photograph of the THGEM electrode. In all measurements there were used electrodes with 400 μm diameter holes, 120 μm rim and 800 μm pitch. The applied technology of the THGEM production provides reproducibility and high precision of the rims that in the long run defines the breaking voltage (maximal gain).

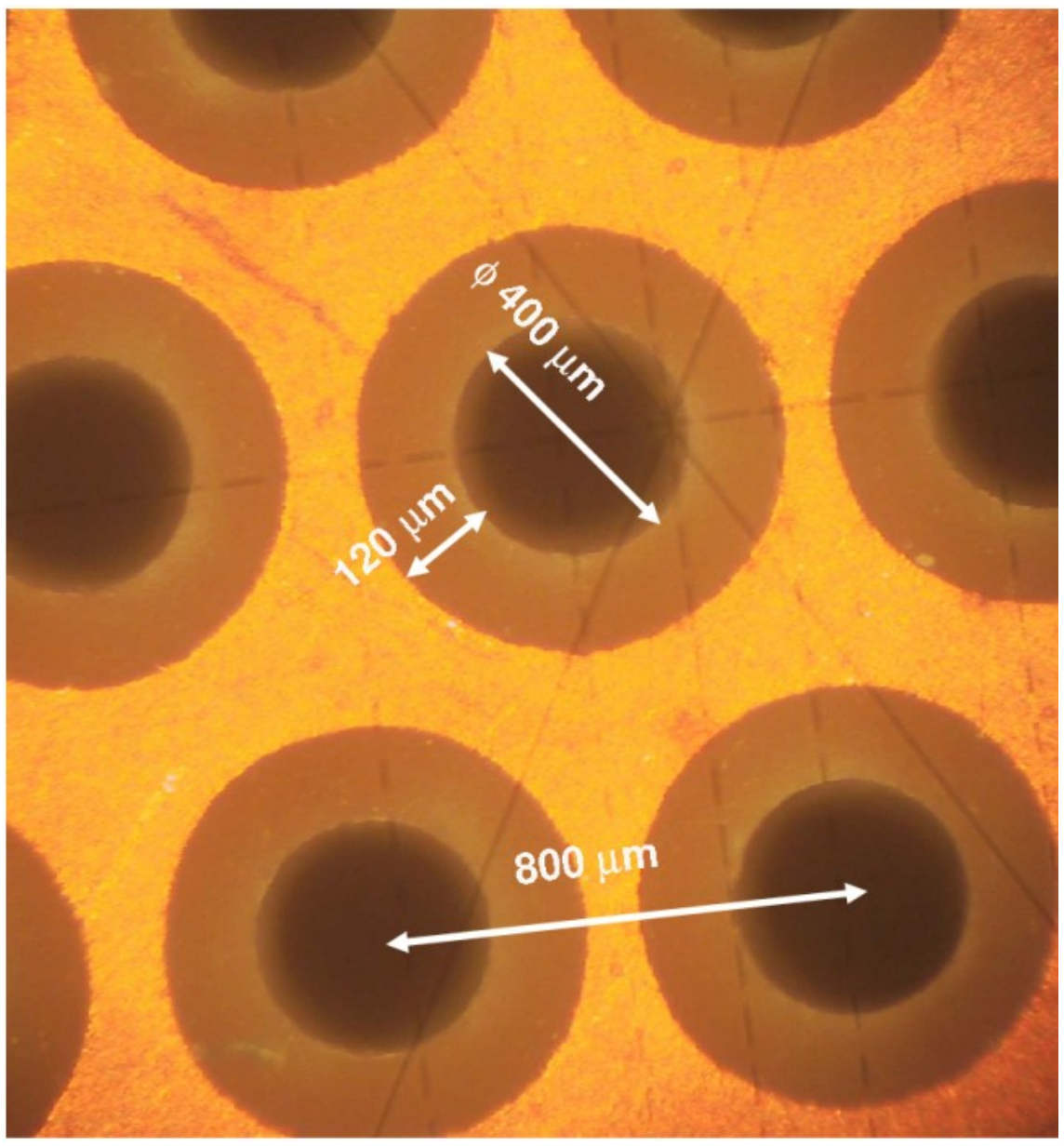

Fig. 3. Microscope photograph of the THGEM electrode.

The measurements presented below were carried out with scheme shown in fig. 1, with drift gap 3 mm, induction gap 1 mm. The detector was flashed continuously with gas mixture of Ar : $CO_2$ (70:30) at atmospheric pressure.

Such design provides the maximum gain about $10^5$ according to the current studies. In many cases it would be desirable to have some reserve. The further increase of the gain can be achieved by cascaded scheme presented in fig. 4. The device is more complicated in construction then the single THGEM. To simplify the design – to eliminate the gas gap between the THGEMs -- the two THGEM were joined as shown in fig 5. The three layers THGEM was produced using the same technology that was used for two layers THGEM that is drilling and etching of the rims were done only once. The rim of the inner layer was the same as of the outer layers. The photo of the electrode is shown in fig. 6. The electrode was inserted between the high voltage and signal electrodes. The chamber was operated at the same conditions as the previous one. High voltage was fed through a 10 MΩ resistor to each electrode using separate sources.

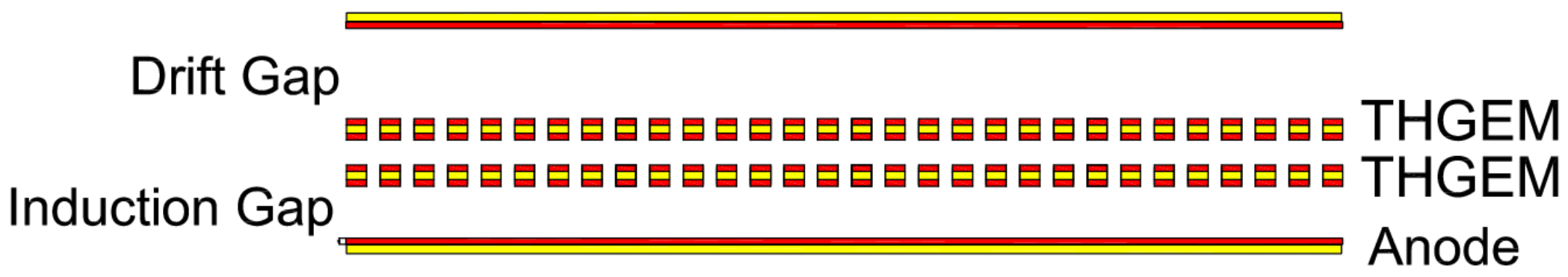

Fig. 4. Double THGEM detector scheme.

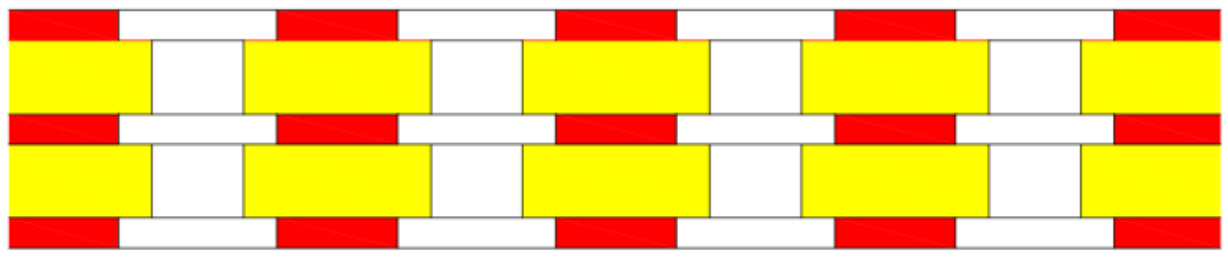

Fig. 5. Fragment of the three layers THGEM electrode.

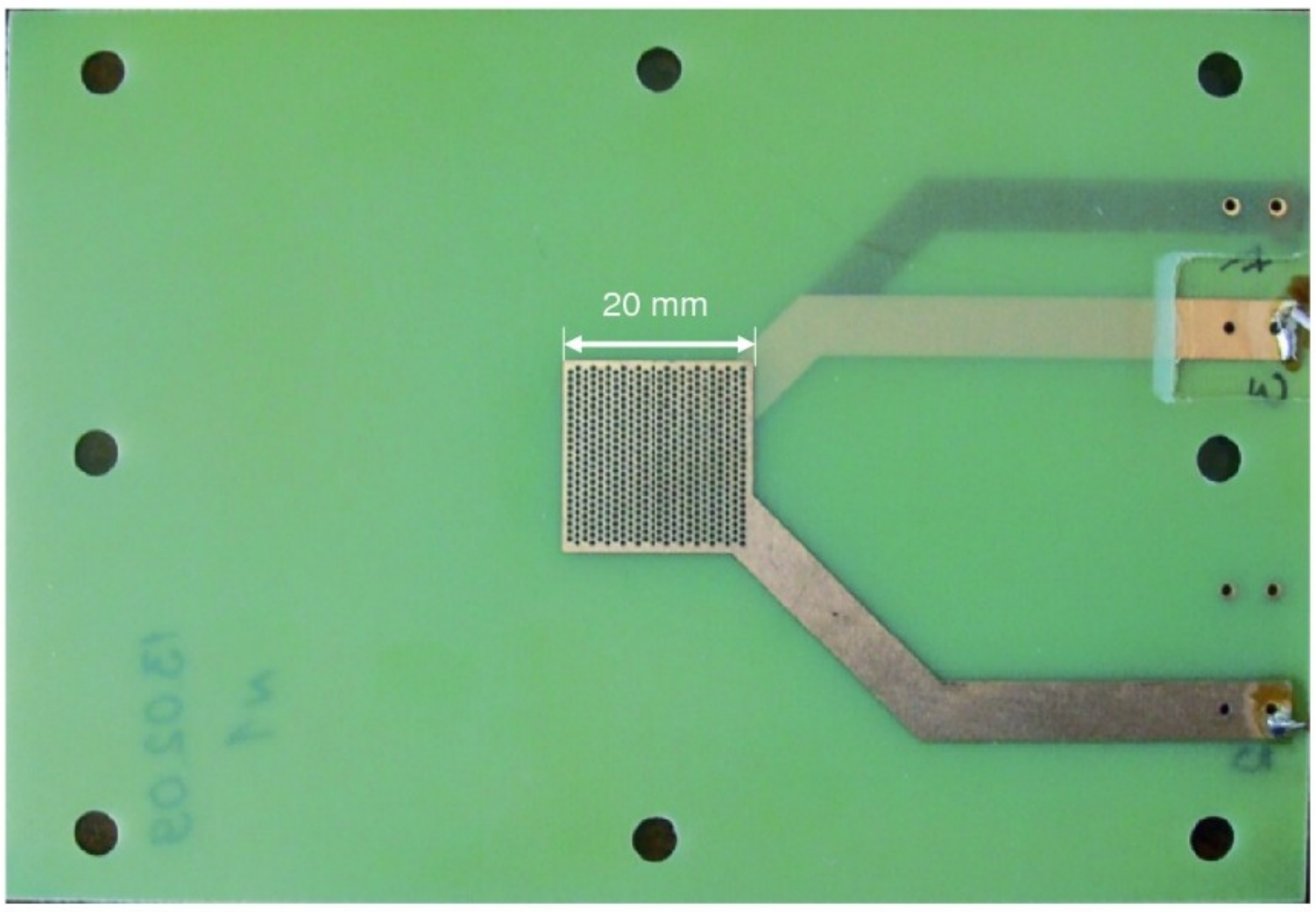

Fig. 6. Photo of the three layers THGEM electrode.

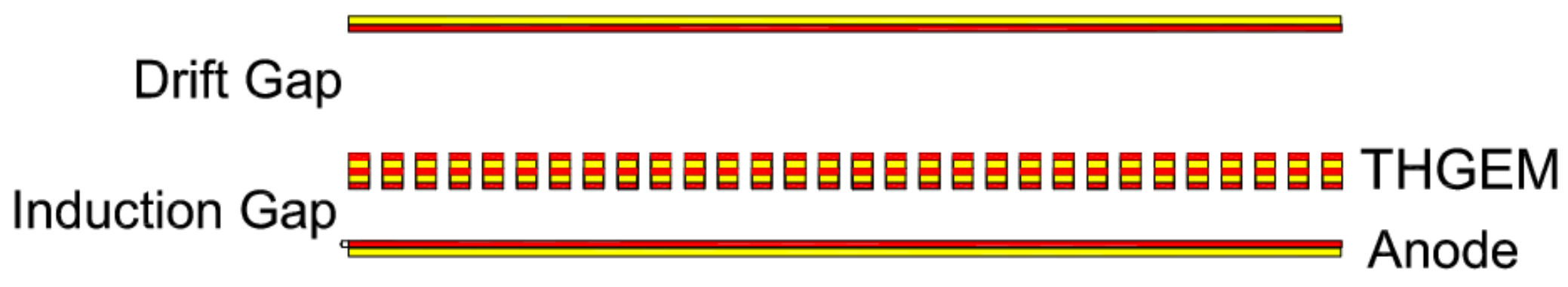

Fig. 7. The three layers THGEM.

The thickness of the electrode was 1 mm. If the cathode and anode plates have thickness 1 mm (to serve as the outer protective covers) then the total thickness of the chamber will be 7 mm and it can be easily inserted into existing absorber of CMS end cap calorimeters.

3. Results

The chamber was irradiated by collimated $^{90}Sr$. Two layers THGEM shown in fig. 2 was operated at the following voltages: $V_{drift}$=600 V, $V_{thgem}$=1.9 kV and $V_{induct}$=200 V. The pulse amplitude was about 1 mV on the 50 Ω load for electrodes produced at CERN (1 $cm^2$) and at IHEP ($4cm^2$). Above this $V_{thgem}$ the gas breakdown starts. The variation of the $V_{drift}$ and $V_{induct}$ had minor influence.

Fig. 8 shows an example of scope pulse from the three layers THGEM presented in fig 7. In this case $V_{drift}$=600 V, $V_{thgem1}$=1.9 kV, $V_{thgem2}$=1.5 kV and $V_{induct}$=200 V. The signal from the anode was amplified and fed to multichannel analyzer. Fig. 9 presents the pulse-height distribution for this chamber irradiated by $^{90}Sr$ without trigger, pedestal in zero channel. In the first approximation such distribution will correspond to minimum ionization particle (muon). The estimation of the number of the primary electrons by the analyzes of the distribution gives the value ~13.

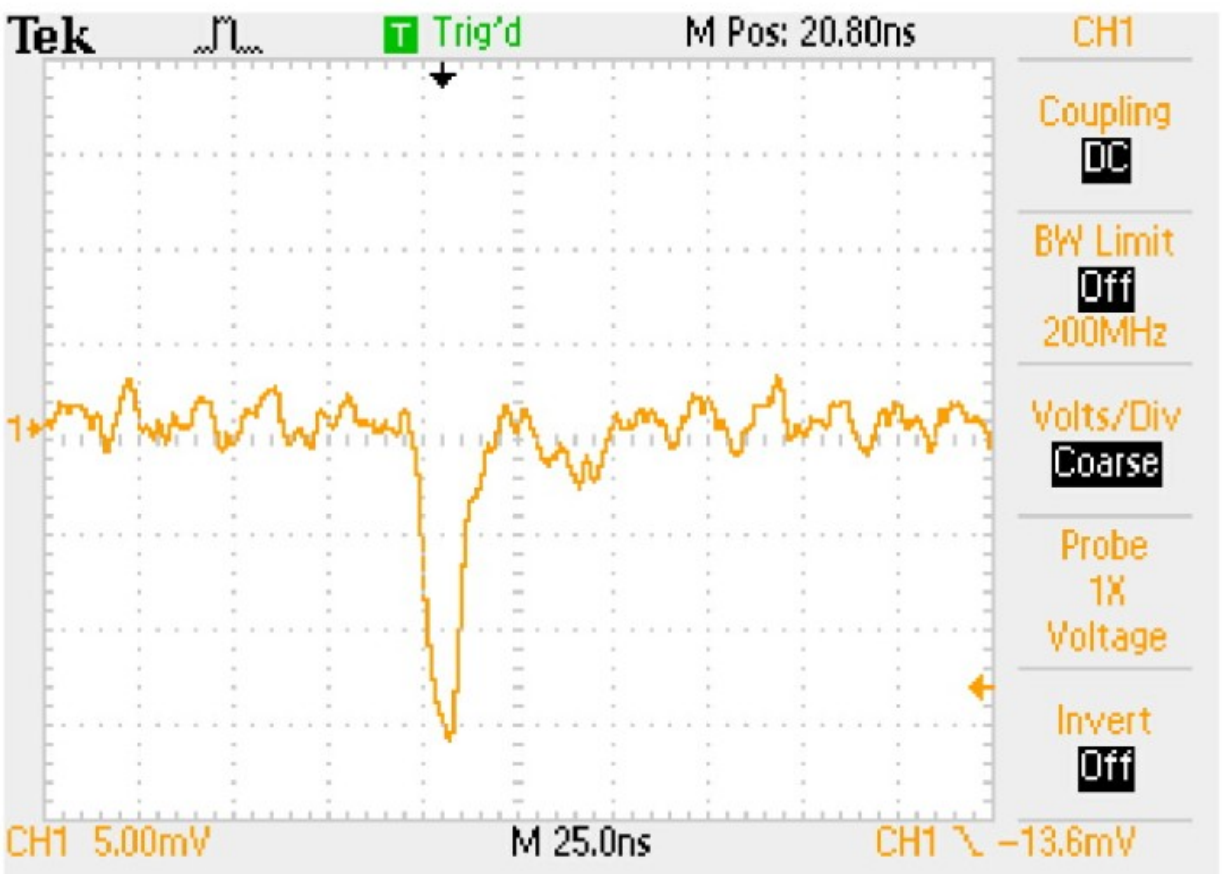

Fig. 8. A scope pulse from detector shown in fig. 7 irradiated with $^{90}Sr$.

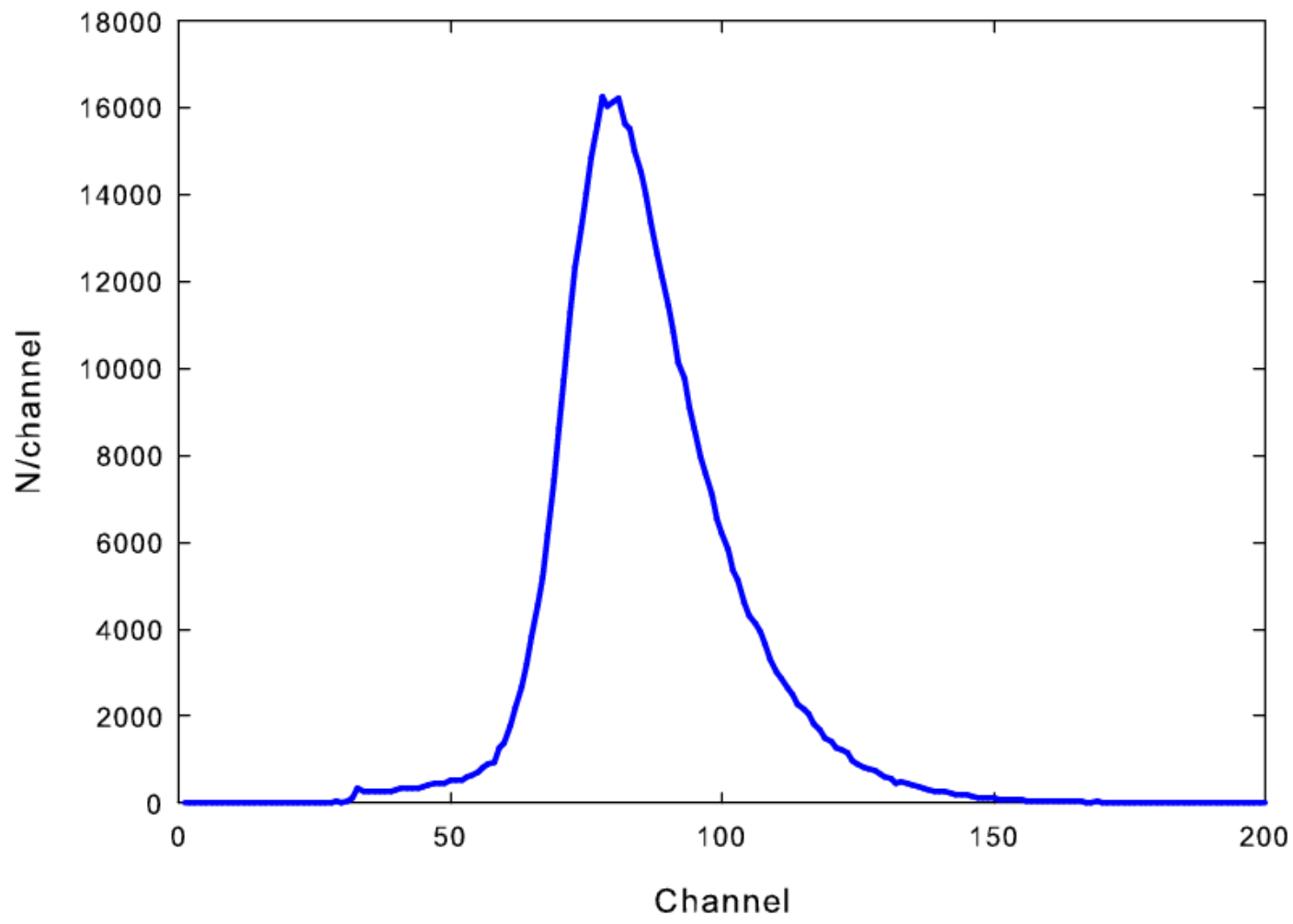

Fig. 9. Pulse-height spectra recorded by multichannel analyzer without trigger with radioactive source $^{90}$Sr.

For comparison fig. 10 shows pulse-height distribution for 300 GeV muons from endcap hadron calorimeters of CMS [5].

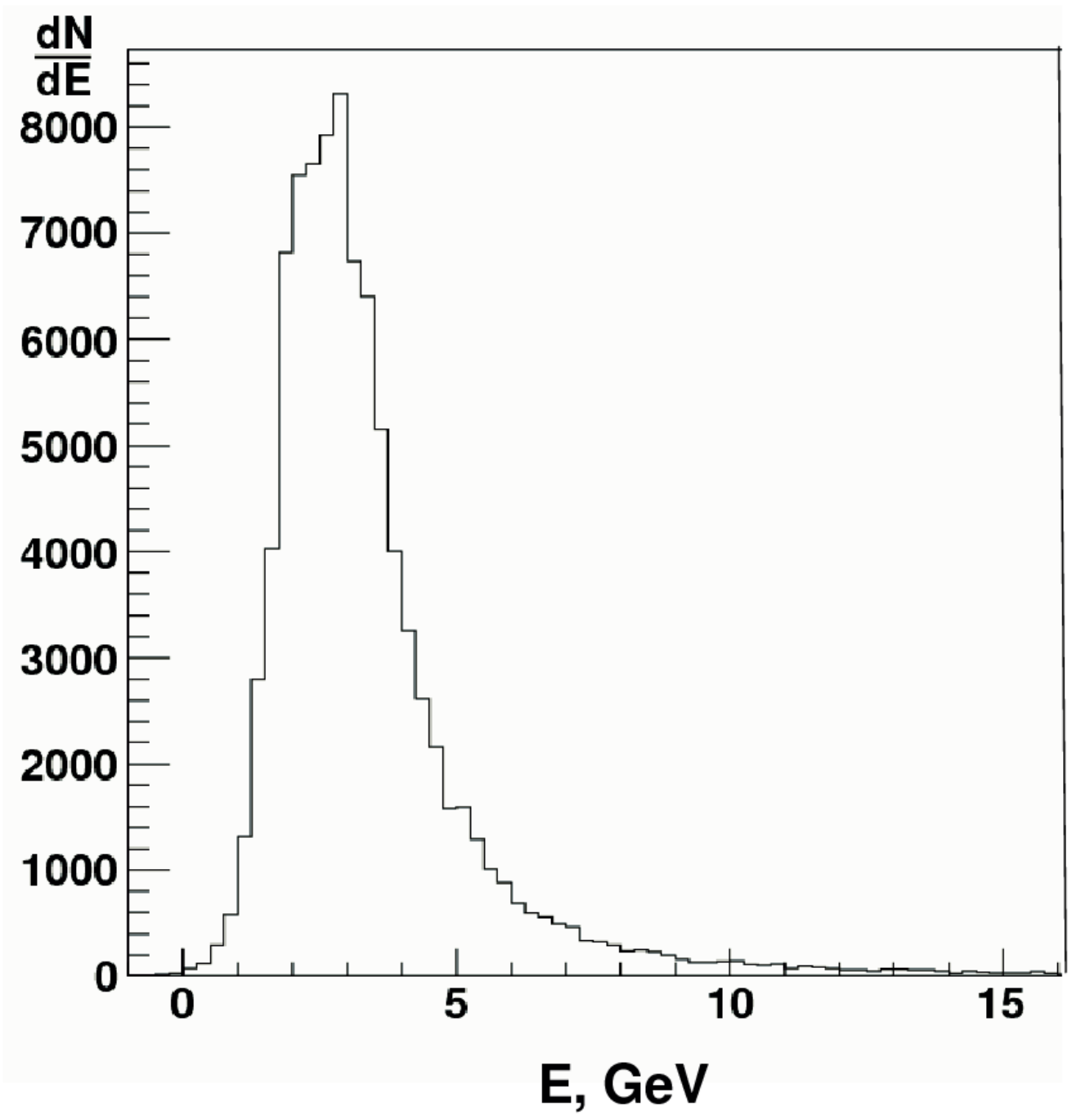

Fig. 10. Pulse-height spectra of 300 GeV muons recorded by CMS end cap hadron calorimeter (18 layers of scintillators). The pedestal is in zero channel.

There was also produced four layers THGEM displayed in fig. 11. The gain increased an order of magnitude in comparison with the three layers THGEM. The main disadvantage of the design is the rise of the high voltage (about 2 kV). Besides the maximum

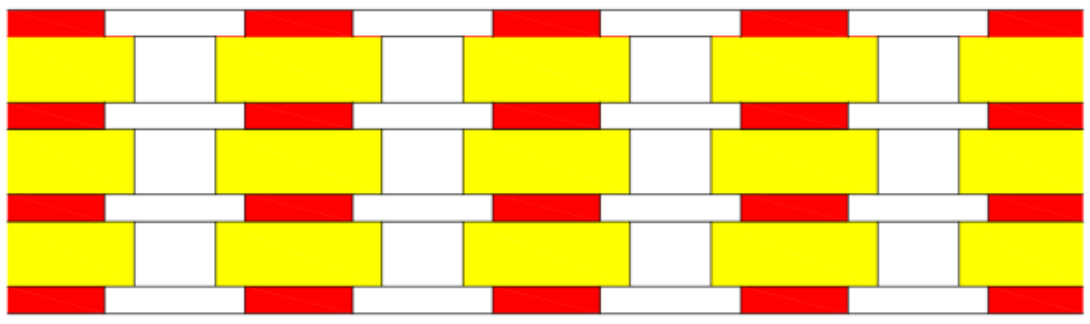

Fig. 11. Fragment of the four layers THGEM.

achievable charge is determined by the so-called Raether limit [6] :

$$An_0 < 10^8 \text{ electrons},$$

where $A$ is the gain, , $n_0$ is the number of the primary ionization electrons. For stable performance this value must be lower by one or two order of magnitude. The used design of the electrode obviously has lower gain in comparison with divided electrodes. But what is the gain for the calorimeters we must strive for? As one can see in fig. 10 the maximum distribution corresponds to 3 GeV. If we set THGEV voltage to 1 mV pulse amplitude for mip in one layer then the total pulse height for mip will be 18 mV. Correspondingly for 300 GeV hadron (or jet) the pulse amplitude will be 100 times higher – 1.8 V. Taking into account that present photodetectors used for end cap hadron calorimeters have gain about $10^4$ and collect ~ 3 photoelectrons from a layer the 3 layers THGEM can be considered acceptable one.

To simplify the design further (to do away with induction gap) there was also studied scheme presented in fig. 12. This is so-called CAT [8] or WELL [9]. Primary advantage of this approach is the robustness of the structure but in this case the gain drops substantially in accordance with other observations [9]. Taking into account the evident complication of the production this direction was not persuaded further also.

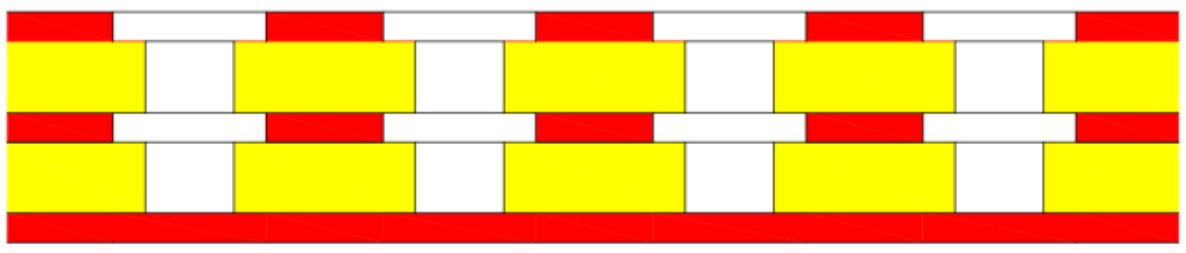

Fig. 12. Fragment of the two layers CAT or WELL.

4. Conclusion.

The THGEM optimization study (gain/robustness) allowed to start the design of the calorimeter prototype we hope to test at the end of the year. We plan also to produce small chambers to place them in strong magnetic field of the spectrometer FODS [10] to study long term performance of the detectors at high counting rate. One of this chamber will be irradiated to 60 Mrad at IHEP radiation center to study its radiation hardness.

Acknowledgements

One of us (N.K) thanks G. Britvich for technical help.